# Quantifying the coupling between strain and cation valence in high entropy oxide thin films using electron microscopy

*Sai Venkata Gayathri Ayyagari[1], Saeed SI Almishal[1], Debangshu Mukherjee[2], Kevin M. Roccapriore[3,4], Jon-Paul Maria[1], and Nasim Alem[1*]*

[1.] Department of Materials Science and Engineering, The Pennsylvania State University, University Park, Pennsylvania, United States.

[2.] Computational Sciences & Engineering Division, Oak Ridge National Laboratory, Oak Ridge, Tennessee, United States.

[3.] Center for Nanophase Materials Sciences, Oak Ridge National Laboratory, Oak Ridge, Tennessee, United States.

[4.] AtomQ, Knoxville, Tennessee, United States.

* Corresponding author: nua10@psu.edu

## Abstract

High entropy oxides (HEOs) are a class of materials with vast compositional space and tunable properties, making them attractive for applications in thermoelectrics, magnetism, ionic conduction, and beyond. However, their metastable nature makes the local structure, and consequently their properties, highly sensitive to growth conditions. It is therefore essential to probe the local modulations in atomic, chemical, and electronic structure as a function of growth conditions. Here, advanced S/TEM techniques, including 4D-STEM combined with electron energy loss spectroscopy and energy-dispersive X-ray spectroscopy are used to investigate the effect of substrate temperature on structure and strain at the nanoscale regime in HEO thin films. We quantify how nanoscale strain variations correlate with Co valence and subtle chemical differences in the films with the same nominal composition but different growth temperatures. Our results demonstrate that identical HEO compositions can accommodate distinct strain and defect states in thin film form and highlight how synthesis conditions can be leveraged to manipulate strain and Co valence. These findings establish a framework to tailor functional properties via strain and valence control in high entropy oxide thin films.

## 1. Introduction

High entropy oxides (HEOs) are multicomponent oxides in which several cations share a single crystallographic sublattice, generating large configurational entropy, local lattice strain, and often metastable structures with a wide range of tunable properties.[1–5] Functional properties like magnetism are highly sensitive to local lattice distortions, cation charge states, defect chemistry, making HEOs a promising yet complex platform for designing functional materials.[6,7]

The prototypical rock salt HEO ($Mg_{1/5}Ni_{1/5}Co_{1/5}Cu_{1/5}Zn_{1/5}$)O, often referred to as J14, has been shown to exhibit tunable structural, electronic, and magnetic properties through composition and growth conditions.[6,8–11] Recent electron energy loss spectroscopy (EELS) and 4D STEM measurements on J14 thin films has further shown that changes in growth conditions can drive substantial changes in cation valence and strain state, which may be tensile or compressive depending on growth temperature.[10] These thin films exhibit modulations in magnetic behavior and optical band gap, underscoring the need to understand the underlying local electronic and strain landscapes. [8] This raises a central question: can we tune and quantify the local strain further by modulating the cation valence through deliberate changes in composition and growth conditions? Such nanoscale strain and local valence modulations can be effectively characterized using advanced S/TEM techniques such as 4D-STEM and STEM-EELS.

There are growing efforts to control strain effects in HEOs by leveraging lattice distortion arising from the presence of multiple elements,[12] manipulating growth conditions,[10,13] or changing the substrate to induce different strain states.[14] In addition, aliovalent cation substitution can itself induce local lattice strain due to size and charge mismatch, making $Sc^{3+}$ incorporation in J14 an attractive pathway to tune both strain and charge balance. $Sc^{3+}$ bears a higher charge than the divalent cations in J14, with an ionic radius similar to high spin $Co^{2+}$ but slightly larger than those of the other cations in J14, thereby introducing charge imbalance and local distortions. These perturbations must be compensated through changes in transition-metal valence, cation vacancies, and/or oxygen non-stoichiometry, which can be probed by combining EELS and EDS measurements at the nanoscale. It has been shown that non-equilibrium techniques such as pulsed laser deposition (PLD) can introduce high growth temperatures and incorporation of cations with significant size or charge mismatch, such as Sc, Ca, Cr, and Ga, into the J14 lattice.[15–17] Previous work has also shown that the structure and stability of Sc-alloyed J14 depend strongly on oxygen partial pressure and growth rate, highlighting the importance of synthesis conditions for controlling the local environment of Sc and the associated defect chemistry.[18] These modulations are the key to finely tuning the local strain and, in turn, a wide range of properties.

In this work, we focus on the influence of Sc incorporation in J14 on local structure, cation valence, and nanoscale strain in PLD grown six-component J14+Sc thin films, using advanced S/TEM. Emerging 4D-STEM approaches, especially when combined with unsupervised machine learning, now enable quantitative mapping of local strain and structural heterogeneities in complex oxide thin films at the nanometer scale, providing a powerful lens on HEO disorder.[19,20] This work will focus on J14+Sc grown at two different growth temperatures to assess the effects of growth kinetics on the atomic, chemical structure, and the local strain in the J14+Sc system.

We begin by examining the reciprocal space using 4D STEM and determination of the strain variations in J14+Sc films grown at two different temperatures. EELS is then employed to probe electronic changes as a function of growth temperature, revealing spatial variations in Co valence states that correlate with strain variations. The atomic and chemical structure probed by STEM-EDS indicate the overall compositional homogeneity of the films and a relatively higher concentration of cation vacancies in the film grown at lower temperature (300 °C) compared to the film grown at higher temperature (500 °C), indicating charge compensation through vacancy formation. These results demonstrate that aliovalent Sc substitution combined with growth temperature control provides an effective strategy to manipulate strain and Co valence in J14+Sc HEOs, establishing a framework for strain- and valence-engineered HEO thin films. They also demonstrate the critical role of advanced S/TEM in resolving coupled strain, chemical, and electronic variations that are inaccessible by bulk characterization alone.

## 2. Results and Discussion

### 2.1. Overview of sample and 4D STEM experiment

In this study, a high entropy oxide thin film with the composition $(Mg_{1/6}Ni_{1/6}Co_{1/6}Cu_{1/6}Zn_{1/6}Sc_{1/6})O$, hereafter referred to as J14Sc (referred to elsewhere as JSc), was grown at two different substrate temperatures: 300°C (lower temperature, J14Sc1) stacked on top of 500°C (higher temperature, J14Sc2), on an MgO substrate, as shown in **Figure 1a**. This sample configuration allows a direct comparison of the influence of growth temperature on the local chemistry, electronic structure, and lattice strain within the same sample system. Magnetic exchange bias results shown in **Figure 1b**, reveal changes in the exchange bias response between J14 and J14Sc thin films, both grown at 500 °C, confirming that the magnetism can be tuned by aliovalent substitution in high entropy oxide systems. Similarly, growth kinetics can have a profound impact on the resulting atomic and chemical structure and, consequently, its macroscale properties.

To elucidate the influence of growth kinetics on structural nuances in J14Sc thin films, reciprocal space analysis is performed via selected area electron diffraction (SAED) and 4D STEM. SAED across the stacked films (shown in **Figure S1**) confirms rock salt structure with no evidence of secondary phases. A tetragonal distortion of the rock salt phase is observed by a split in the out-of-plane (004) reflection, while the in-plane $(0\bar{4}0)$ reflection remains sharp, as shown in the magnified image insets in **Figure S1**. From the electron diffraction along the [100] zone, we observe approximately a 1.6% increase in the out-of-plane lattice parameter of J14Sc thin film grown at 500°C (J14Sc2) compared to J14Sc thin film grown at 300°C (J14Sc1). A similar increase in the out-of-plane lattice parameter was observed by our collaborators in macroscopic X-ray diffraction measurements.[21]

To quantify strain variations across the thin film stack, we performed 4D STEM experiments. 4D-STEM is a technique in which a converged probe is rastered across a two-dimensional space, and at each probe position, a two-dimensional reciprocal-space pattern is acquired.**[19]** Any change in the real space distances appears as an inverse change in reciprocal space (as shown schematically in **Figure 1c**), which can be used to quantify the strain. This technique is considered sophisticated and highly accurate for strain quantification and can be used to measure strain variations over several tens of nanometers even under local sample tilts.**[22]** Geometric Phase Analysis (GPA) is another method that quantifies the phase variation in Fourier space arising from translational variations in real space. The phase variations along non-collinear directions are then used to quantify strain; however, this method is highly sensitive to the accuracy of the chosen reciprocal lattice vectors.**[23,24]**

Multiple 4D-STEM datasets were acquired across the entire stack of thin films and MgO substrate. The average converged beam electron diffraction (CBED) patterns from two distinct regions (Region 1 and Region 2) of the same FIB lamella are displayed in **Figure 1d and 1e**, respectively. As shown in **Figure 1a**, two boxed regions schematically indicate the areas from which datasets were collected in the same FIB lamella. Both regions span the thin film stack and substrate, and were collected from different locations in the same FIB lamella. The simultaneously acquired ADF-STEM image from Region 1 and Region 2 are shown in **Figure 2a** and **Figure S2c**, respectively.

The average CBED from Region 2 (**Figure 1e**) shows additional reflections (marked with circles) that are absent in Region 1. These additional reflections do not match spinel reflections previously observed in J14 and related systems at certain growth conditions.[11,16,17] Additional reflections can indicate the presence of secondary phases or ordering in the material, specifically in the case of high entropy oxide systems with the presence of multiple cations and metastable nature. Further analysis of 4D STEM data in Region 2 reveals that additional reflections observed in some regions of the TEM sample arise from local sample bending and confirms that there is no local ordering or secondary phase formation in the J14Sc thin films at this growth conditions, as shown in **Figure S2**.

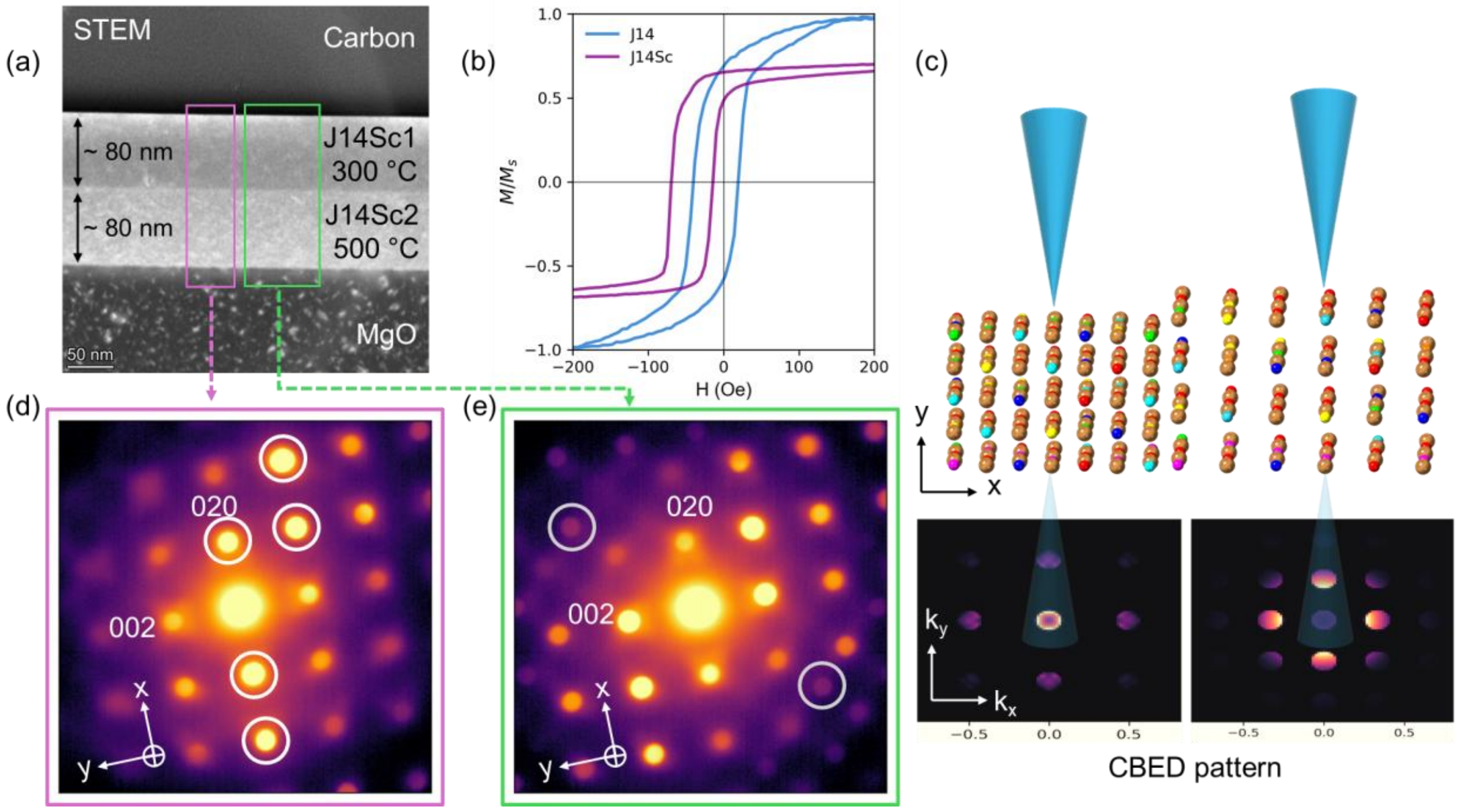

**Figure 1. Motivation and overview of the 4D STEM experiment:** (a) LAADF STEM image of the cross-section showing the thin film stack (~ 80 nm each), where J14Sc1 is the thin film grown at 300 °C and J14Sc2 is the thin film grown at 500 °C. (b) Magnetic exchange bias results of J14 and J14Sc grown at 500 °C, adapted from references.[8,21] (c) Schematic of 4D

STEM experiment illustrating the strain variations in real space manifest as inverse trends in reciprocal space. (d,e) Average CBED patterns from two different regions of the sample (referred to as Region 1 (pink box) and Region 2 (green box), for brevity). (d) Circular insets indicate the reflections used to perform strain mapping in **Figure 2**. (e) Circular insets highlight additional reflections observed in Region 2 that are typically absent in the rock salt structure along [100] zone axis. These reflections originate from sample bending, as discussed in **Figure S2**.

### 2.2 Strain mapping using 4D STEM

To determine the strain in the film, we have analyzed Region 1 (region with no bending; STEM image of this region is shown in **Figure 2a**). The 4D-STEM strain mapping is performed on the entire film stack. A subtle tilt is observed when going from the top thin film to the MgO, as observed in **Figure S3**. The summed CBED of MgO is used as the reference. CBED disks that are well cross-correlated are chosen for the 4D-STEM strain mapping, as shown in **Figures S4** and **S5**.

The strain map results are shown in **Figures 2**. The strain along the y axis (out-of-plane direction) $\boldsymbol{\varepsilon_{yy}}$ exhibits a strong tensile strain in both thin films. The extent of tensile strain in J14Sc1 grown at 300 °C is different from that in J14Sc2 grown at 500 °C. There is an increase in the out-of-plane lattice strain in the J14Sc film grown at 500 °C compared to J14Sc grown at 300 °C. On average, the tensile strain in J14Sc1 (top film) is approximately 1.42% relative to the MgO substrate, while J14Sc2 (bottom film) has approximately 3.11% tensile strain relative to the substrate. The average strain across each row of the thin film stack is plotted in **Figure S6**.

Strain map along x axis (in-plane direction) $\boldsymbol{\varepsilon_{xx}}$ shown in **Figure 2d** shows that differences in in-plane strain (with respect to MgO substrate) are relatively small and in compressive state compared to the out-of-plane component. We also observe that although the extent of strain change in $\boldsymbol{\varepsilon_{xx}}$ (~ 0.42% in top film and ~ 0.26% in bottom film) is not as large as in $\boldsymbol{\varepsilon_{yy}}$, there are still subtle strain variations. Furthermore, J14Sc2 near the interface between J14Sc1 and J14Sc2 exhibits a strong change in strain in $\boldsymbol{\varepsilon_{xx}}$, $\boldsymbol{\varepsilon_{xy}}$, $\boldsymbol{\varepsilon_{\theta}}$.

To understand the projected volumetric change ($\boldsymbol{\varepsilon_{vv}}$) in the unit cell, $\boldsymbol{\varepsilon_{xx}}$ and $\boldsymbol{\varepsilon_{yy}}$ are summed and plotted in **Figure S6**, highlighting changes in the overall projected unit cell of J14Sc grown at different temperatures, predominantly dominated by strain changes in the out-of-plane (growth direction). The average $\boldsymbol{\varepsilon_{vv}}$ is 1% for the thin film grown at lower substrate temperature and 2.85% for the thin film grown at the higher substrate temperature.

To ensure the strain trends are consistent across different TEM samples, GPA was performed on the TEM specimen used for oxidation state measurements. As shown in **Figure S7**, the J14Sc2 thin film grown at 500 °C exhibits an increased tensile strain compared to the J14Sc1 top thin film grown at 300 °C, which is consistent with the results observed from 4D STEM experiments. The 4D-STEM results confirmed our previous observation from the SAED and further quantified the large-scale strain profile.

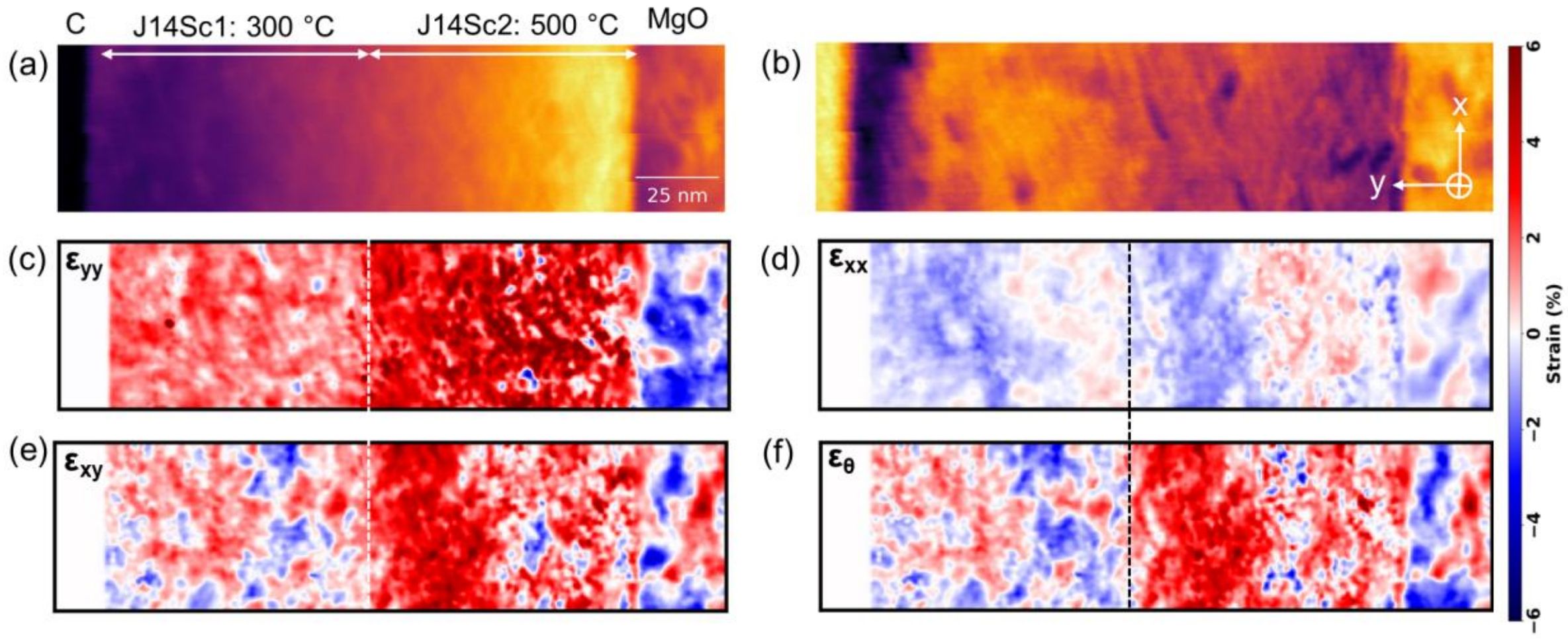

**Figure 2. Increase in out-of-plane tensile strain in films grown at 500 °C vs 300 °C.** (a) Simultaneously acquired ADF-STEM image from Region 1. (b) Virtual BF-STEM image from Region 1. (c-f) Strain maps of Region 1 $\varepsilon_{yy}$, $\varepsilon_{xx}$, $\varepsilon_{xy}$, $\varepsilon_{\Theta}$, respectively with dashed lines indicating approximately the interface of the two thin films.

### 2.3 Electronic Structure and Valence State Analysis

To understand the origin of strain variation in J14Sc thin films grown at different substrate temperatures, we investigate the electronic structure and probe possible changes in cation valence states using spatially resolved electron energy loss spectroscopy (EELS). A subtle change in the pre-edge of O K-edge is observed in the J14Sc thin film grown at 300 °C compared to the one grown at 500 °C, as illustrated in **Figure S8**. The Co L-edge also exhibits slight changes in peak shape, whereas the Ni and Cu L-edges remain consistent throughout the stacked films.

To better understand the changes in the Co L-edge, we performed monochromated EELS and plotted the signal from each row in **Figure 3b**; the corresponding probe positions are shown in **Figure 3a**. A gradual change in the Co L-edge peak shape is clearly visible when moving from the 300 °C grown film to the 500 °C grown film. To further interpret these differences, we summed the monochromated EELS data from each thin-film region, as shown in **Figure 3c**, and compared them to reference Co L-edge spectra from CoO ($Co^{2+}$) and $Co_3O_4$ (mixed $Co^{2+}$/$Co^{3+}$).[25] The Co L-edge from the 500 °C layer closely matches the CoO ($Co^{2+}$) reference, while the 300 °C layer resembles a mixture of $Co^{2+}$ and $Co^{3+}$.

To quantify these valence states, we performed least-squares fitting using the reference spectra (**Figures S10c and S10d**). We find that Co in J14Sc grown at 300 °C consists of 56.75% CoO and 43.25% $Co_3O_4$, corresponding to an average Co valence of approximately 2.289 + (~2.3+). In contrast, the Co in J14Sc grown at 500 °C consists of 95.66% CoO and 4.34% $Co_3O_4$, yielding an average Co valence of approximately 2.029 + (~2+). These findings point out that the presence of Sc in the system limits the tendency of Co to adopt a higher valence state. In a previous study of J14, Co was observed to adopt a valence of 2.6+ at lower temperature (200 °C) and 2.1+ at higher temperature (500 °C).[8,10]

To explore the changes in the O K edge, we similarly summed the spectra from each region, as shown in **Figure S9c**. A subtle difference is observed near 530.8 eV. Since the pre-peak of O K edge arises from the hybridization of O 2p orbitals with transition metal 3d states,[26,27] this variation further supports the observed changes in Co valence. Finally, we examined the average valence states of other cations in the film, as shown in **Figure S10**. We observe that Cu and Ni remain predominantly in the 2+ state, while Sc is in the 3+ state.

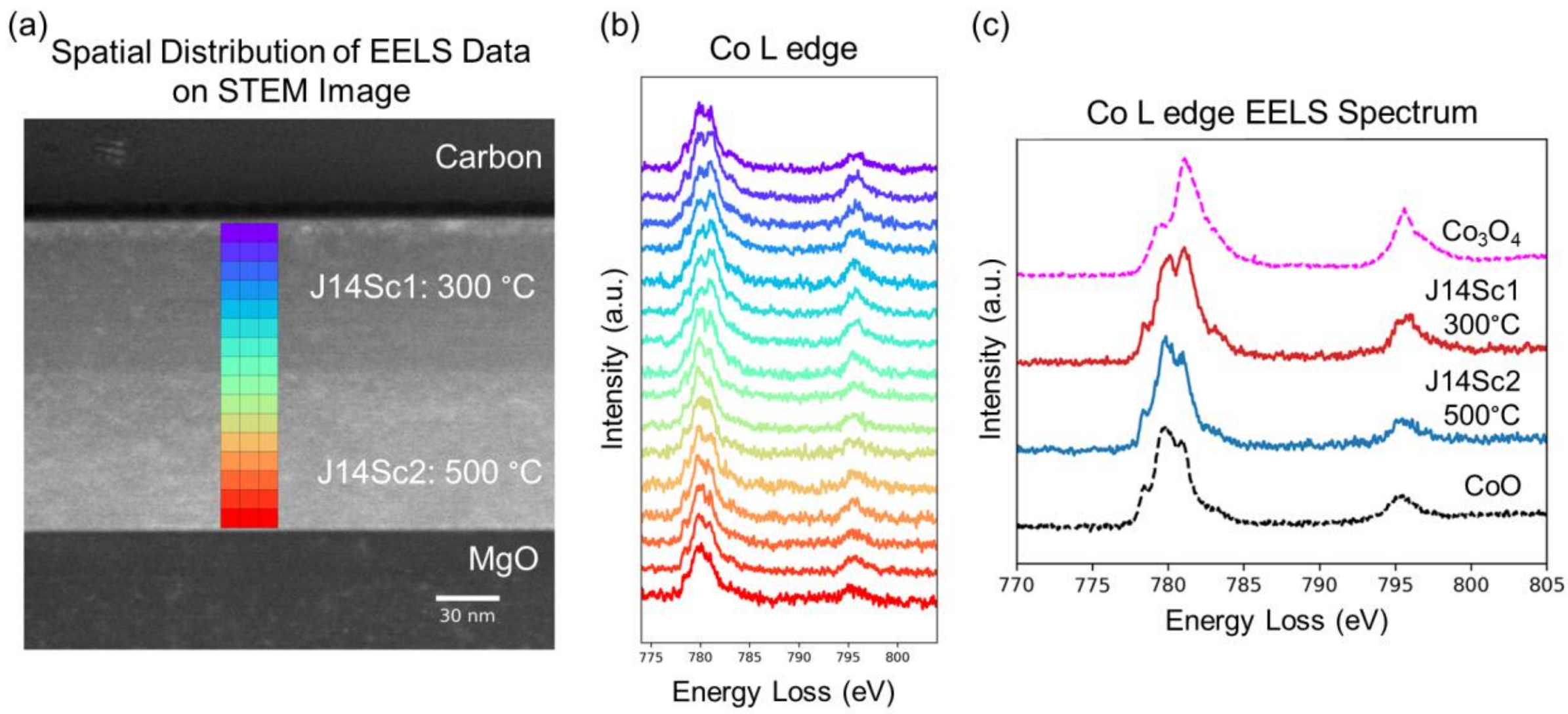

**Figure 3. Increase in out-of-plane strain correlated with decrease in Co valence.** (a) ADF-STEM image acquired simultaneously with the EELS spectrum, with colors corresponding to the EELS data shown in (b). (b) Co L edge from each row across the stacked thin film. (c)

Average Co L-edge spectra from J14Sc1 and J14Sc2, plotted alongside $Co^{2+}$ (CoO) and $Co^{2+/3+}$ ($Co_3O_4$) reference EELS data.[25]

## 2.4 Local Structure and Chemical Investigation with STEM and EDS

To understand how the observed strain and valence changes influence the structure and chemistry, we investigated the local structure and chemical nuances of the stacked films using STEM imaging and EDS mapping. The HAADF-STEM image of the cross-section (**Figure 4a**) clearly shows the interface between the J14Sc1 thin film grown at 300 °C and the J14Sc2 film grown at 500 °C. To quantify structural differences, we applied a sliding window fast Fourier transform (FFT) across the image, followed by dimensionality reduction using principal component analysis (PCA) and clustering via a Gaussian mixture model (GMM), as shown in **Figure 4b**. The resulting cluster map identifies two structurally distinct regions that align with the two growth temperatures. Insets in **Figures 4b** show the average FFTs for J14Sc1 (Cluster 1, mapped in red color) and J14Sc2 (Cluster 2, mapped in blue color), respectively. The FFT's match with the rock salt structure along [100] axis. A comparison of these FFTs highlights a shift in the out-of-plane reflections, consistent with a change in the lattice parameter. This difference is further emphasized in the FFT difference map (FFT in the middle of **Figure 4b**). Finally, atomic resolution STEM images of J14Sc1 and J14Sc2 (insets in **Figures 4a**), also confirm that both layers maintain a rock salt structure, with a slight change in out-of-plane lattice parameter reinforcing the presence of growth temperature dependent strain effects confirming the absence of any ordering in this system.

STEM-EDS mapping across the stacked film shows chemical homogeneity at this scale, as illustrated in **Figure 4c**. To better understand the local changes in elemental distribution, we summed the EDS intensity from each row of the map and plotted the results in **Figure S12**. To account for local thickness variations when analyzing intensity changes, we normalized the cation intensities with respect to oxygen for each row, as shown in **Figure 4d**. We observe that cations such as Co, Ni, and Cu exhibit a clear decrease in intensity in the thin film grown at 300 °C. This suggests the presence of a higher concentration of cation vacancies in the film grown at the lower temperature (300 °C) compared to the one grown at the higher temperature (500 °C). However, a reverse trend is observed for Mg and Zn. Interestingly, there is a dip in cation intensity at the interface of the stacked films, where all cations except Cu show reduced intensity. Similar behavior was observed at the interface of J14 stacked thin films grown at 200°C and 500°C.[10]

To maintain charge neutrality in the presence of increased Co oxidation, we propose that the J14Sc film grown at 300 °C contains a higher concentration of cation vacancies. This hypothesis is consistent with the EDS analysis presented in **Figure 4d**, indicating a higher concentration of cation vacancies in thin film grown at 300 °C. This STEM-EDS study is an initial support for anamolous crystal growth in high entropy oxide systems.[21]

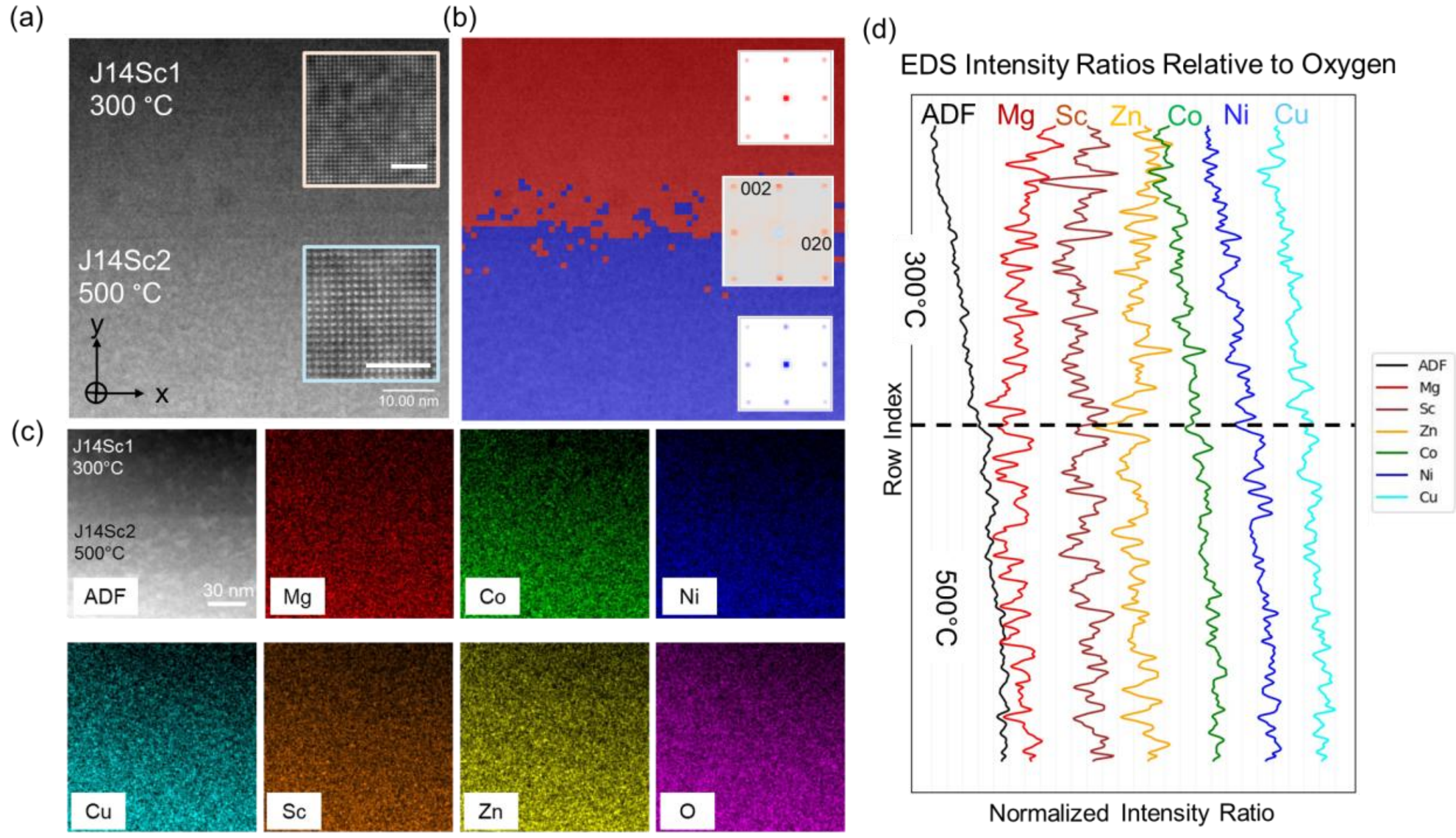

**Figure 4. Cation content decreases at lower growth temperatures to adapt for charge neutrality while maintaining rock salt structure.** (a) HAADF-STEM image of the cross-section of the stacked films, where J14Sc1 is the film grown at 300 °C and J14Sc2 is the film grown at 500 °C, with insets showing the magnified images of each thin film (scalebar is 2 nm). (b) Corresponding cluster map obtained by applying principal component analysis (PCA) followed by Gaussian mixture model (GMM) clustering on sliding window FFTs, overlaid with average FFT from each cluster, red corresponding to J14Sc1 and blue corresponding to J14Sc2 and the difference between J14Sc1 and J14Sc2 FFTs in the middle. (c) EDS maps across the stacked thin films. (d) Intensity plot of EDS maps relative to oxygen. The intensity from each row of the EDS map is summed and plotted with respect to the oxygen intensity.

Our results show the change in the Co valence observed by STEM-EELS correlates with the strain variations in the thin films quantified using 4D-STEM. In J14Sc grown at 300 °C, Co tends to adopt a mixed valence of 2+ and 3+. According to Shannon ionic radii, the ionic radius of $Co^{3+}$ in octahedral coordination (0.545 Å for low spin and 0.61 Å for high spin) is smaller than that of $Co^{2+}$ in octahedral coordination (0.65 Å for low spin and 0.745 Å for high spin). This indicates a reduction of almost 27% when comparing low-spin $Co^{3+}$ to high-spin $Co^{2+}$ [28] and explains why J14Sc grown at 300 °C with mixed Co valence shows a lower tensile strain (smaller lattice parameter) than J14Sc grown at higher temperature.

Changes in the strain state of J14 thin films grown at different temperatures were previously reported.[10] In J14, the film grown at lower temperature exhibits compressive strain, while the film grown at higher temperature exhibits tensile strain relative to MgO. In contrast, in J14Sc, tensile strain is observed at both low and high growth temperatures relative to MgO, which is attributed to the presence of Sc. Sc is a larger 3+ cation with an ionic radius similar to high-spin $Co^{2+}$and about 7.97% larger than Ni,[28] the smallest cation in J14. So on average J14Sc has a larger effective cation radius than J14 and therefore a larger lattice parameter. As noted earlier, the presence of Sc also limits the ability of Co to adopt higher valence states compared to J14, further contributing to the larger lattice parameter. Since Co valence states and strain in the structure influences the magnetic behavior, we expect to see a change in magentic behavior for J14Sc grown at different temperatures, similar to J14.[8]

This work shows that the combined effects of composition (i.e., the presence of Sc) and growth temperature influence both the chemical environment and the strain in HEO systems. Together, these observations demonstrate that coupled control of composition (via Sc substitution) and growth temperature enables simultaneous tuning of strain and cation valences in HEO thin films, providing a pathway to engineer their functional properties like magnetism.

## 3. Conclusion

This work demonstrates a direct correlation between local lattice strain, electronic and, chemical environment in J14Sc HEO thin films using a combination of advanced electron microscopy techniques 4D STEM with STEM-EELS and EDS. J14Sc films exhibit tensile strain of approximately 1.42% and 3.11% relative to the MgO substrate in the out-of-plane direction at low (300 °C) and high (500 °C) growth temperatures, respectively. At low growth temperatures, a mixed $Co^{2+}/Co^{3+}$ valence state and a relatively higher concentration of cation vacancies modify the strain state compared to the film grown at higher temperature. The larger ionic radius of $Sc^{3+}$ relative to most other cations in J14 leads to an overall expansion of the lattice in J14Sc compared to J14. These results show that Sc substitution and growth temperature can be used as effective knobs to tune both strain and cation valence states in HEO thin films, suggesting a general strategy for engineering functional properties such as magnetism among others. Strain engineering, already established as a central strategy in oxide research, thus offers a natural framework to expand the functional landscape of high entropy oxides. The combination of compositional complexity and epitaxial strain provides a promising route to stabilize new phases, enhance functional properties, and extend the application space of HEOs.

## 4. Experimental Methods

*Material Synthesis:*

The High Entropy Oxide $(Mg_{1/6}Ni_{1/6}Co_{1/6}Cu_{1/6}Zn_{1/6}Sc_{1/6})O$ was synthesized by mixing the binary oxide components (MgO (≥99%), CoO (≥99%), NiO (99.99%), CuO (99.99%), $Sc_2O_3$ (99.99%), and ZnO (99.999%), milling, and pressing them into pellets, followed by sintering in air at 1000°C for 12 h and subsequently air-quenched. These pellets were then used to grow thin films on [001]-MgO substrates via pulsed laser deposition using a 248 nm KrF laser with a fluence of 1.2 J/cm$^2$ at 50mTorr $O_2$ partial pressure. The first layer, adjacent to MgO, was grown at a substrate temperature of 500 °C. After 2000 pulses (~ 80 nm of film thickness, as confirmed in **Figure 1a**), the substrate heater was set to 300 °C. Once the temperature stabilized, another 2000 laser pulses were applied, resulting in an additional~ 80 nm of film growth, as confirmed by cross-sectional TEM in **Figure 1a**.

*TEM Sample Preparation:*

Sample preparation for Scanning/Transmission Electron Microscopy (S/TEM) studies were carried out using FEI Helios 660 Dual Beam Focused Ion Beam (FIB). The TEM lamellae were extracted along [100] zone and thinned at 30kV and 5kV ion beam, respectively. The final cleaning was performed at 2kV ion beam to minimize the damage caused by $Ga^+$ ions. Multiple FIB samples were extracted to ensure if we are capturing the overall behavior of these stacked thin films.

*Scanning/Transmission Electron Microscopy:*

The STEM experiments were performed at an accelerating voltage of 300 kV using a double Cs-corrected FEI Titan G2 microscope equipped with a monochromator. The LAADF STEM image shown in **Figure 1a** was acquired at a collection angle of 27–64 mrad. The HAADF STEM image was acquired at a collection angle of 63–368 mrad. Strain quantification using geometric phase analysis was conducted on the STEM image with Strain++ software.[23]

*EDS analysis:*

The EDS intensity plots were obtained by applying a Gaussian blur with a 5×5 pixel kernel, and the intensity sums of each row were normalized and plotted together in **Figure S12a**. To counteract the effect of local thickness variations, the intensity of each cation was plotted relative to the intensity of oxygen and then normalized.

*EELS analysis:*

The electron energy loss spectroscopy is obtained in monochromated set up with the energy resolution of ~0.3 eV at a dispersion of 0.025 eV/channel. The low loss and core loss spectra are simultaneously acquired. The processing of EELS data including zero loss peak alignment and the EELS $t/\lambda$ map estimation is performed in Python using Hyperspy.[29] To enhance the signal to noise ratio, the EELS data is denoised using singular value decomposition (SVD) for data denoising through dimensionality reduction. Non-negative least squares fitting of the EELS data using reference spectra were performed in Python to estimate the average valence.

*4D STEM data acquisition:*

The 4D STEM datasets were acquired using the Nion UltraSTEM U100 at Oak Ridge National Laboratory. The microscope was operated at an accelerating voltage of 100 kV with a convergence semi-angle of 2.5 mrad to prevent overlap of diffraction disks. The scan step size varied between 0.5 nm (for the strain mapping dataset) and 2.5 nm (for bend contour analysis dataset). The 4D STEM dataset used for strain mapping was acquired over 100 x 400 grid in real space, with each probe position capturing a CBED pattern sampled on a 300 × 300-pixel grid in reciprocal space. Because an inherent relative rotation exists in the CBED data, the axes were marked accordingly.

The 4D STEM datasets were collected from a different sample than the one used for STEM imaging and STEM- EELS data acquisition. GPA measurements were carried out on the same sample as the EELS dataset to confirm consistency of the results.

For analysis of additional reflections and bend contours, a separate dataset (80 × 80 positions in real space) was collected. All data processing was performed in Python using custom scripts that utilized toolkits such as STEMTooL, Py4DSTEM and other standard packages like NumPy, scikit-learn, sciPy.[30,31] Virtual dark field images were generated by integrating the intensity of selected reflections across each probe position in the 4D STEM dataset, thereby enabling spatial mapping of specific reciprocal space features.

*Strain Mapping using 4D STEM Data:*

The strain was quantified by comparing the stacked thin film unit cell with a reference cell at each scanning pixel. To accurately determine the CBED disk center, the data were preconditioned by applying a logarithmic transformation followed by Sobel filtering to enhance the disk edges, as described in the literature.[24] The preprocessed CBED pattern was then cross-correlated with a template diffraction disk obtained from the amorphous carbon region (**Figure**

**S4**). This process converts the disk location into a sharp peak, which was fitted with a two-dimensional Gaussian function to identify the peak position with subpixel precision in diffraction space (**Figure S5**). Disks that exhibited strong cross-correlation were selected for precise strain measurement. According to previous studies, strain mapping remains effective even with slight sample tilt, provided that non-collinear diffraction disks are chosen for the analysis.[22]

## Acknowledgements

The authors acknowledge support from the Penn State Materials Research Science and Engineering Center for Nanoscale Science under National Science Foundation award DMR-2011839. The authors also acknowledge use of the Penn State Materials Characterization Lab. The 4D STEM experiments and analysis were conducted as part of a user project (CNMS2024-A-02275) at the Center for Nanophase Materials Sciences (CNMS), a U.S. Department of Energy, Office of Science User Facility at Oak Ridge National Laboratory. D.M. was funded by the INTERSECT Initiative of the Laboratory Directed Research and Development (LDRD) Program of Oak Ridge National Laboratory (ORNL). ORNL is managed by UT-Battelle, LLC, for the US Department of Energy under contract DE-AC05-00OR22725.

## References

[1] C. M. Rost, E. Sachet, T. Borman, A. Moballegh, E. C. Dickey, D. Hou, J. L. Jones, S. Curtarolo, J.-P. Maria, *Nat. Commun.* **2015**, *6*, 8485.

[2] G. N. Kotsonis, S. S. I. Almishal, F. Marques dos Santos Vieira, V. H. Crespi, I. Dabo, C. M. Rost, J.-P. Maria, *J. Am. Ceram. Soc.* **2023**, *106*, 5587.

[3] S. S. I. Almishal, J. T. Sivak, G. N. Kotsonis, Y. Tan, M. Furst, D. Srikanth, V. H. Crespi, V. Gopalan, J. T. Heron, L.-Q. Chen, C. M. Rost, S. B. Sinnott, J.-P. Maria, *Acta Mater.* **2024**, *279*, 120289.

[4] S. S. I. Almishal, M. Furst, Y. Tan, J. T. Sivak, G. Bejger, J. Petruska, S. V. G. Ayyagari, D. Srikanth, N. Alem, C. M. Rost, S. B. Sinnott, L.-Q. Chen, J.-P. Maria, *Nat. Commun.* **2025**, *16*, 8211.

[5] S. S. I. Almishal, P. Kezer, J. T. Sivak, Y. Iwabuchi, S. V. G. Ayyagari, S. Sarker, M. Furst, G. Bejger, B. Yang, S. Gelin, N. Alem, I. Dabo, C. M. Rost, S. B. Sinnott, V. Crespi, V. Gopalan, R. Engel-Herbert, J. T. Heron, J.-P. Maria, *Adv. Sci.* **2025**, *12*, e09868.

[6] A. Sarkar, R. Kruk, H. Hahn, *Dalton Trans.* **2021**, *50*, 1973.

[7] A. R. Mazza, E. Skoropata, Y. Sharma, J. Lapano, T. W. Heitmann, B. L. Musico, V. Keppens, Z. Gai, J. W. Freeland, T. R. Charlton, M. Brahlek, A. Moreo, E. Dagotto, T. Z. Ward, *Adv. Sci.* **2022**, *9*, 2200391.

[8] G. N. Kotsonis, P. B. Meisenheimer, L. Miao, J. Roth, B. Wang, P. Shafer, R. Engel-Herbert, N. Alem, J. T. Heron, C. M. Rost, J.-P. Maria, *Phys. Rev. Mater.* **2020**, *4*, 100401.

[9] N. J. Usharani, P. Arivazhagan, T. Thomas, S. S. Bhattacharya, *Mater. Sci. Eng. B* **2022**, *283*, 115847.
[10] L. Miao, J. T. Sivak, G. Kotsonis, J. Ciston, C. L. Ophus, I. Dabo, J.-P. Maria, S. B. Sinnott, N. Alem, *ACS Nano* **2024**, *18*, 14968.
[11] S. S. I. Almishal, L. Miao, Y. Tan, G. N. Kotsonis, J. T. Sivak, N. Alem, L.-Q. Chen, V. H. Crespi, I. Dabo, C. M. Rost, S. B. Sinnott, J.-P. Maria, *J. Am. Ceram. Soc.* **2025**, *108*, e20223.
[12] J. Baek, M. D. Hossain, P. Mukherjee, J. Lee, K. T. Winther, J. Leem, Y. Jiang, W. C. Chueh, M. Bajdich, X. Zheng, *Nat. Commun.* **2023**, *14*, 5936.
[13] L. Luo, J. Ju, Y. Wu, X. Wan, W. Li, Y. Li, H. Jiang, Y. Hu, C. Li, *Adv. Mater.* **2025**, *37*, 2418856.
[14] Z. Zhao, A. K. Jaiswal, D. Wang, V. Wollersen, Z. Xiao, G. Pradhan, F. Celegato, P. Tiberto, M. Szymczak, J. Dabrowa, M. Waqar, D. Fuchs, X. Pan, H. Hahn, R. Kruk, A. Sarkar, *Adv. Sci.* **2023**, *10*, 2304038.
[15] C. M. Rost, Entropy-Stabilized Oxides: Explorations of a Novel Class of Multicomponent Materials, **2016**.
[16] G. E. Niculescu, G. R. Bejger, J. P. Barber, J. T. Wright, S. S. I. Almishal, M. Webb, S. V. G. Ayyagari, J. Maria, N. Alem, J. T. Heron, C. M. Rost, *J. Am. Ceram. Soc.* **2025**, *108*, e20171.
[17] S. V. G. Ayyagari, L. Miao, G. Niculescu, M. Webb, J. Barber, J. Heron, C. M. Rost, N. Alem, *Microsc. Microanal.* **2024**, *30*, ozae044.663.
[18] G. N. Kotsonis, C. M. Rost, D. T. Harris, J.-P. Maria, *MRS Commun.* **2018**, *8*, 1371.
[19] C. Ophus, *Microsc. Microanal.* **2019**, *25*, 563.
[20] T. Yoo, E. Hershkovitz, Y. Yang, F. Da Cruz Gallo, M. V. Manuel, H. Kim, *Npj Comput. Mater.* **2024**, *10*, 223.
[21] S. S. I. Almishal, S. V. G. Ayyagari, A. Pearre, P. Kezer, M. Furst, C. M. Rost, B. Yan, N. Alem, T. Charlton, Z. Mao, J. T. Heron, J.-P. Maria, **2026**, DOI 10.48550/arXiv.2603.00814.
[22] D. Mukherjee, H. Yu, C. Wang, J. Spendelow, D. Cullen, M. Zachman, *Microsc. Microanal.* **2021**, *27*, 1440.
[23] M. J. Hÿtch, **n.d.**
[24] D. Mukherjee, J. T. L. Gamler, S. E. Skrabalak, R. R. Unocic, *ACS Catal.* **2020**, *10*, 5529.
[25] P. Ewels, T. Sikora, V. Serin, C. P. Ewels, L. Lajaunie, *Microsc. Microanal.* **2016**, *22*, 717.
[26] J. H. Seo, J.-H. Kwon, *Nanomaterials* **2023**, *13*, 2767.
[27] F. Frati, M. O. J. Y. Hunault, F. M. F. de Groot, *Chem. Rev.* **2020**, *120*, 4056.
[28] R. D. Shannon, *Acta Crystallogr. A* **1976**, *32*, 751.
[29] F. de la Peña, E. Prestat, J. Lähnemann, V. T. Fauske, P. Burdet, P. Jokubauskas, T. Furnival, C. Francis, M. Nord, T. Ostasevicius, K. E. MacArthur, D. N. Johnstone, M. Sarahan, J. Taillon, T. Aarholt, pquinn-dls, V. Migunov, A. Eljarrat, J. Caron, T. Nemoto, T. Poon, S. Mazzucco, sivborg, actions-user, N. Tappy, N. Cautaerts, S. Somnath, T. Slater, M. Walls, pietsjoh, **2025**, DOI 10.5281/zenodo.14956374.
[30] D. Mukherjee, R. Unocic, *Microsc. Microanal.* **2020**, *26*, 2960.
[31] B. H. Savitzky, S. E. Zeltmann, L. A. Hughes, H. G. Brown, S. Zhao, P. M. Pelz, T. C. Pekin, E. S. Barnard, J. Donohue, L. Rangel DaCosta, E. Kennedy, Y. Xie, M. T. Janish, M. M. Schneider, P. Herring, C. Gopal, A. Anapolsky, R. Dhall, K. C. Bustillo, P. Ercius, M. C. Scott, J. Ciston, A. M. Minor, C. Ophus, *Microsc. Microanal.* **2021**, *27*, 712.

# Supporting Information

**Quantifying the coupling between strain and cation valence in high entropy oxide thin films using electron microscopy**

*Sai Venkata Gayathri Ayyagari[1], Saeed SI Almishal[1], Debangshu Mukherjee[2], Kevin M. Roccapriore[3], Jon-Paul Maria[1], and Nasim Alem[1*]*

[1.] Department of Materials Science and Engineering, The Pennsylvania State University, University Park, Pennsylvania, United States.

[2.] Computational Sciences & Engineering Division, Oak Ridge National Laboratory, Oak Ridge, Tennessee, United States.

[3.] Center for Nanophase Materials Sciences, Oak Ridge National Laboratory, Oak Ridge, Tennessee, United States.

[4.] AtomQ, Knoxville, Tennessee, United States.

* Corresponding author: nua10@psu.edu

## S1: Electron diffraction analysis

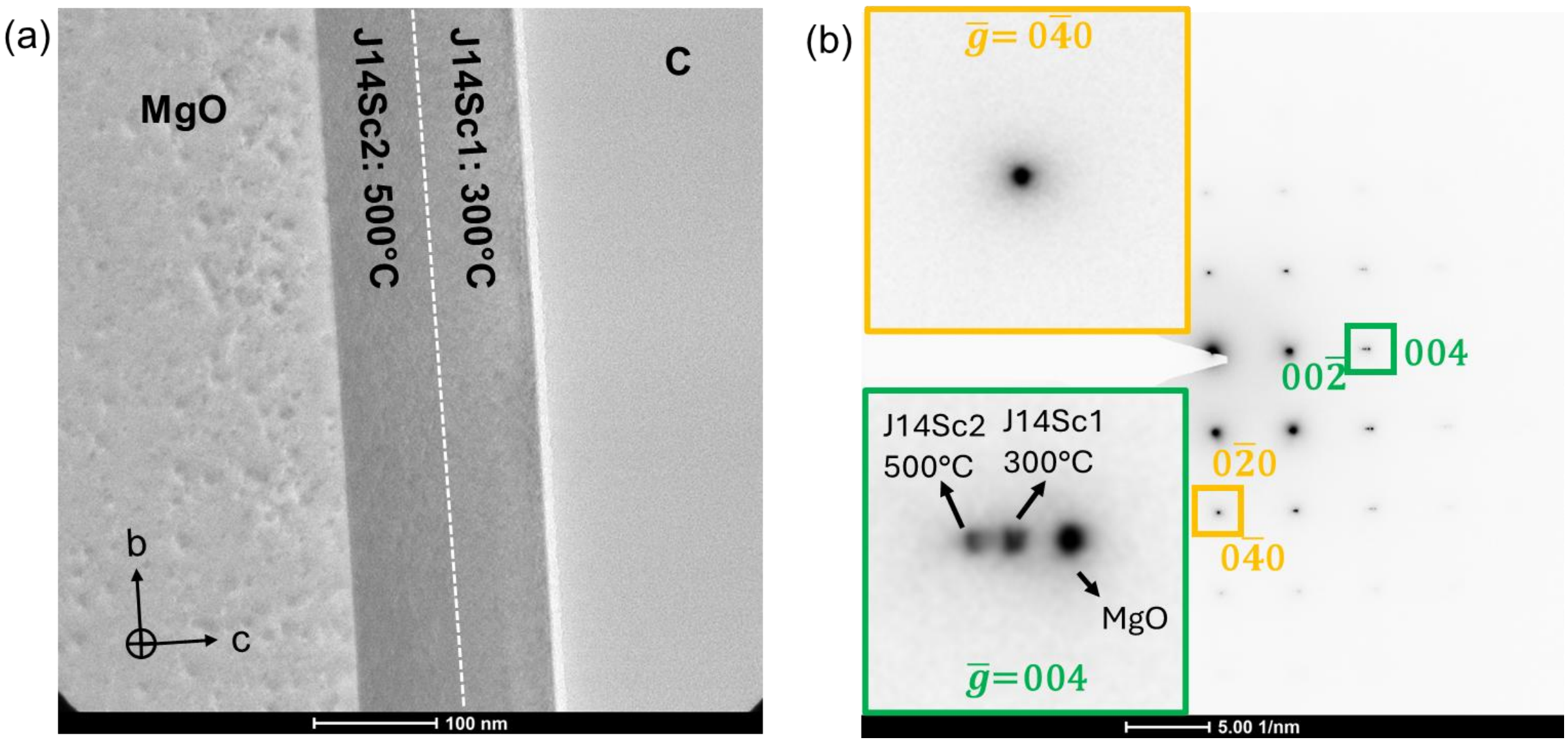

**Figure S1. Selected area electron diffraction analysis.** (a) Selected area corresponding to the selected area electron diffraction (SAED) pattern of the stacked thin films along [100] zone axis in (b). The insets in (b) show the magnified in-plane ($0\bar{4}0$) and out-of-plane (004) reflections.

## S2: Understanding the additional reflections observed in Region 2 of 4D STEM experiments:

To understand the origin of additional reflections in Region 2 (**Figure 1e**), we conducted further analysis to determine if these samples exhibit local ordering missed by initial electron diffraction analyses. Virtual dark field imaging was performed on the 4D STEM dataset (**Figure S2b**), revealing that the additional reflections arise from sample bending. The bent contours giving rise to these additional reflections are also visible in the simultaneously acquired STEM image (**Figure S2c**).

To index the additional reflections, electron diffraction patterns were simulated with slight tilts away from the [100] zone axis (**SI video**) and compared to the experimental data. The simulated patterns matched the additional reflections and the CBED data is indexed as shown in **Figure S2d**, confirming that sample bending induced during preparation causes diffraction from planes typically absent in the ideal [100] zone axis orientation. These findings confirm that the sample does not exhibit any intrinsic ordering or secondary phases; rather, the observed additional reflections result from bending induced during sample preparation. To avoid bias in strain measurements, this dataset was excluded from strain analysis.

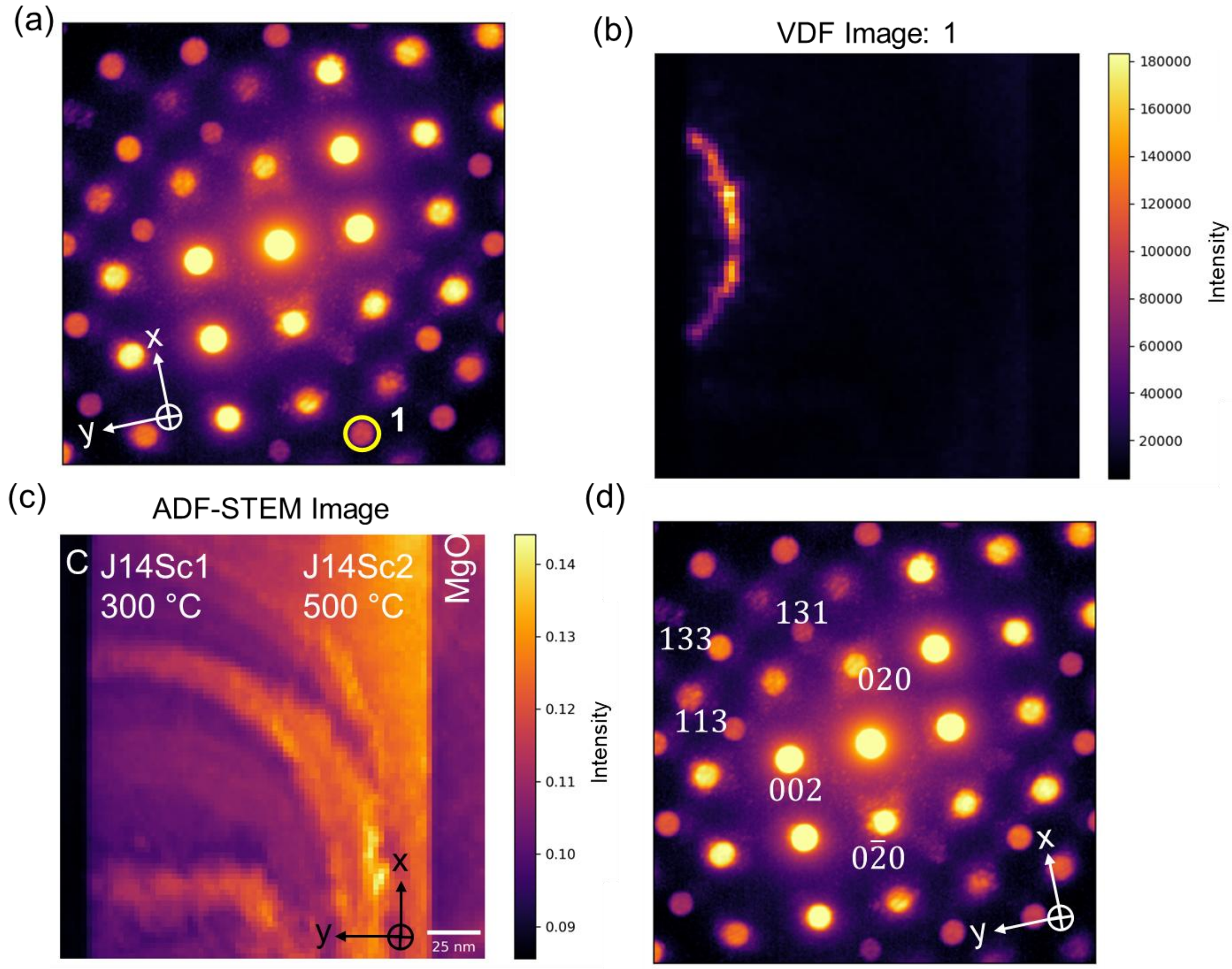

**Figure S2. Sample bending creates additional reflections in 4D STEM.** (a) CBED pattern showing the maximum diffraction intensity from Region 2, with an inset corresponding to the disc used to generate the (b) virtual dark-field image. (c) Simultaneously acquired ADF-STEM image representing the Region 2. (d) Indexed CBED reflections compared with the simulated diffraction pattern shown in the **SI video**. These additional reflections, attributed to sample bending, are especially pronounced in the 4D-STEM dataset due to the detector's high sensitivity and the extensive reciprocal-space sampling.

**S3-S7: Strain mapping using 4D STEM experiment:**

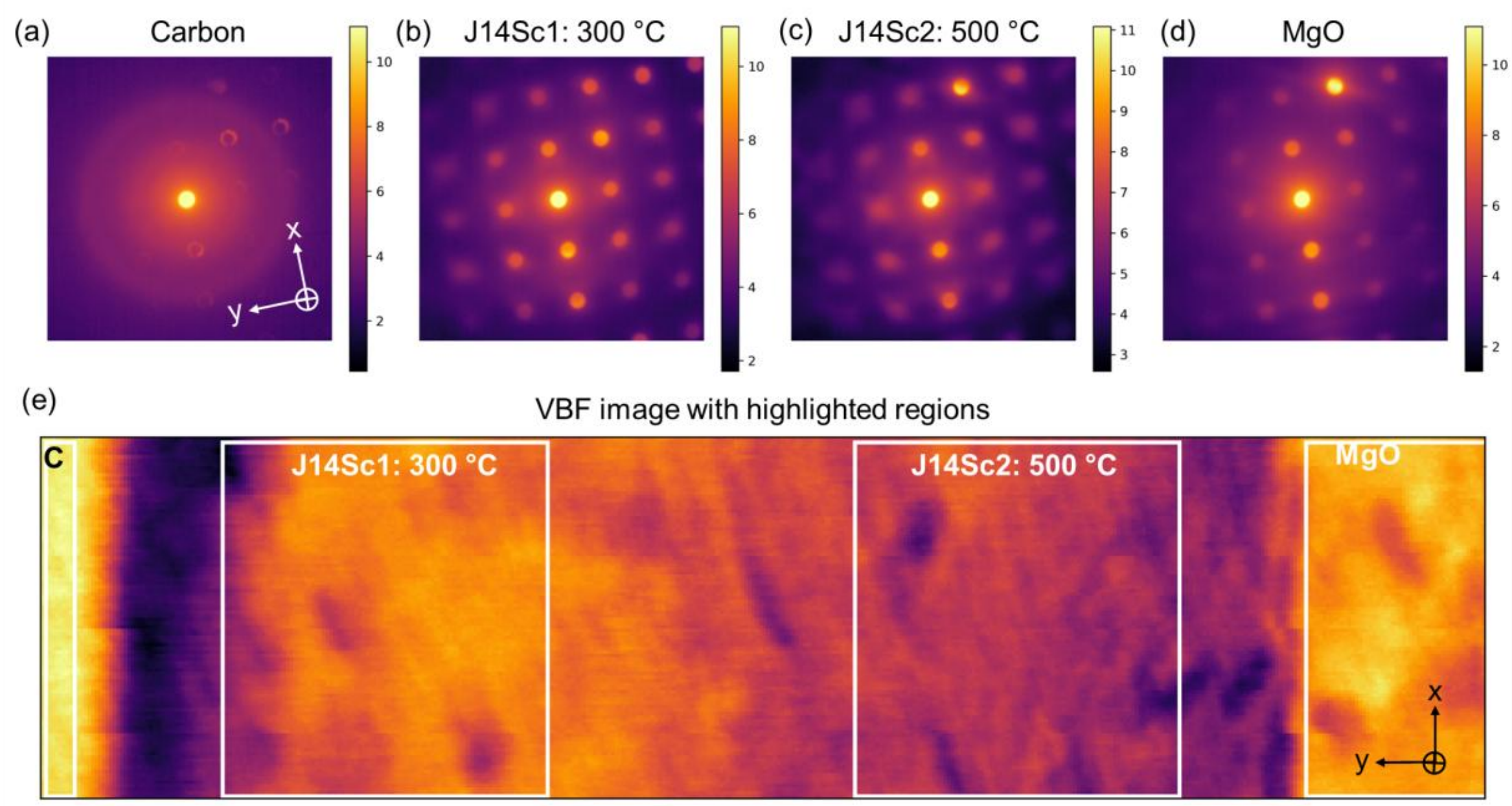

**Figure S3. Summed CBED patterns from different regions.** (a–d) Summed CBED patterns from different regions of the same sample, as shown in the virtual bright-field (VBF) image with highlighted regions in (e). The boxes represent the regions from which the average CBED patterns are obtained.

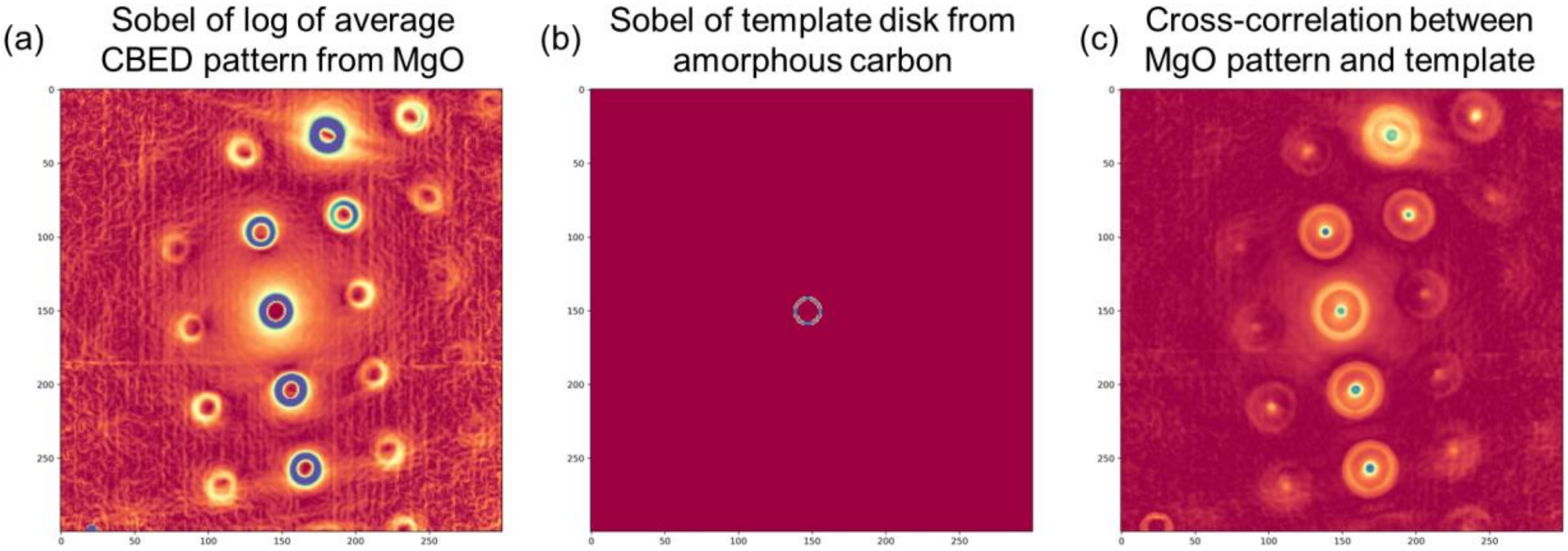

**Figure S4. Sobel filtering and cross-correlation for disk detection.** (a) Sobel-filtered logarithm of the average CBED pattern from MgO, as shown in **Figure S3d**. (b) Template disk from amorphous carbon after Sobel filtering. (c) Cross-correlation map between the MgO CBED pattern and the amorphous carbon template.

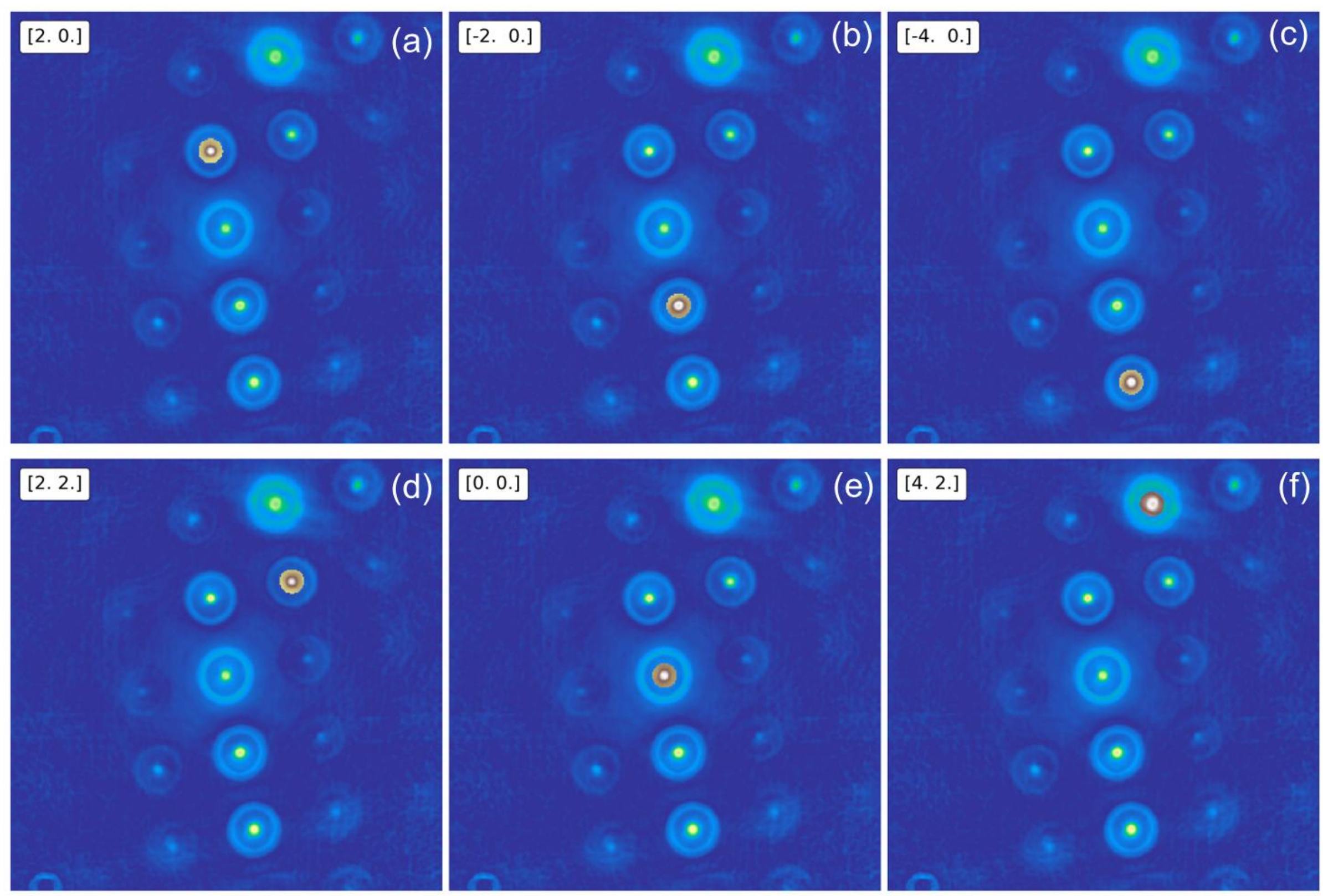

**Figure S5. Diffraction disks used for strain mapping.** (a-f) Diffraction disks indexed on the logarithm of the average CBED pattern from MgO (**Figure S3d**).

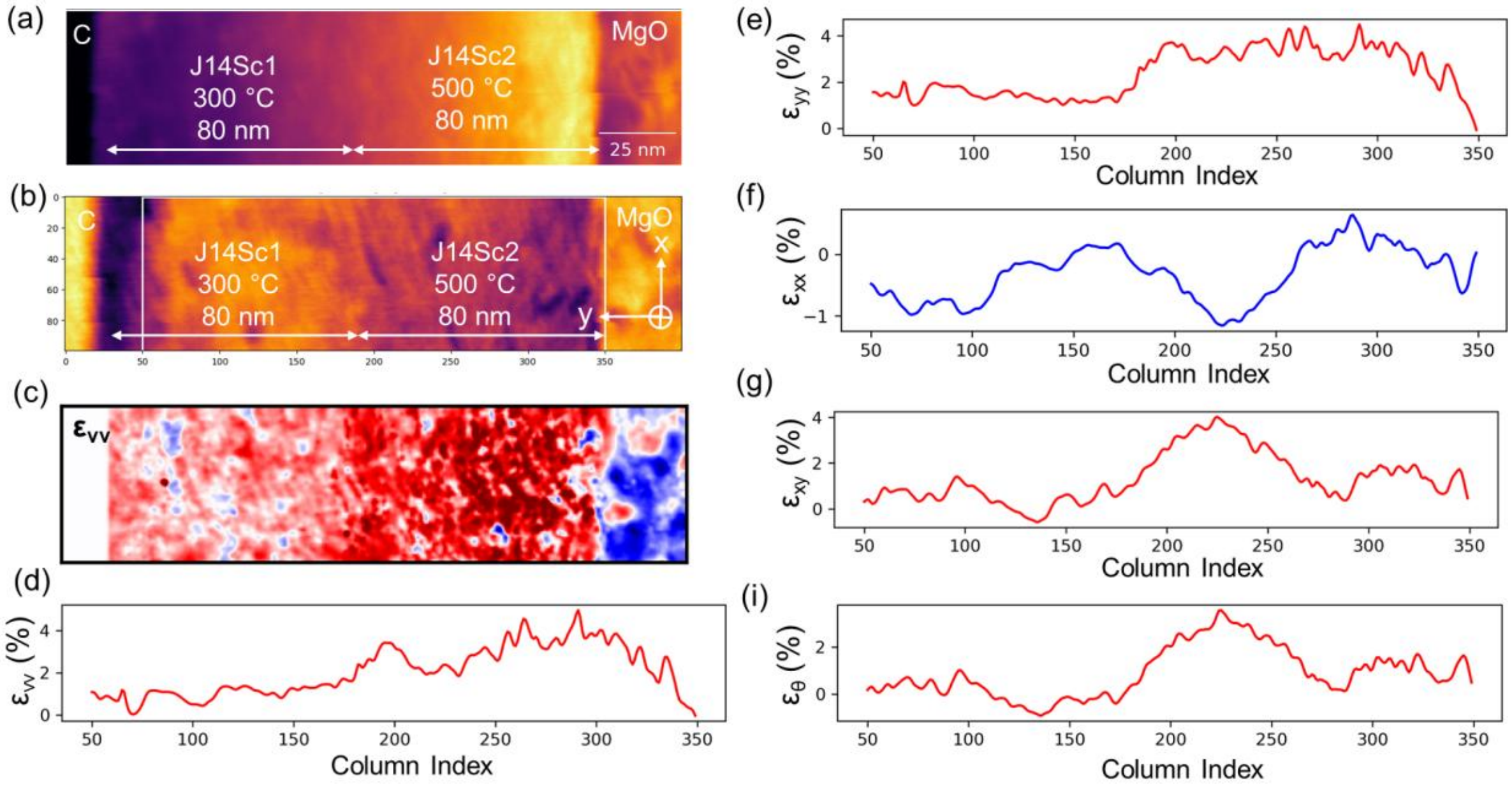

**Figure S6. Calculated strain of the stacked film using 4D-STEM** (a) Simultaneously acquired ADF STEM image. (b) VBF image with a rectangle representing the [50:350] index range, highlighting the region whose strain values are plotted in panels (d–i). (c) Projected volume strain $\varepsilon_{vv}$ calculated as $\varepsilon_{xx} + \varepsilon_{yy}$. To calculate the average strain for each thin-film

dataset, the [50:175] range was used for the top film and [200:350] for the bottom film. The average $\varepsilon_{xx}$ of the top film (J14Sc1 at 300 °C) and bottom film (J14Sc2 at 500 °C) is −0.42% and −0.26%, respectively, where the negative sign indicates compressive strain along the x direction. The corresponding $\varepsilon_{yy}$ values are 1.42% and 3.11%, and the $\varepsilon_{vv}$ values are approximately 1% and 2.85%, respectively.

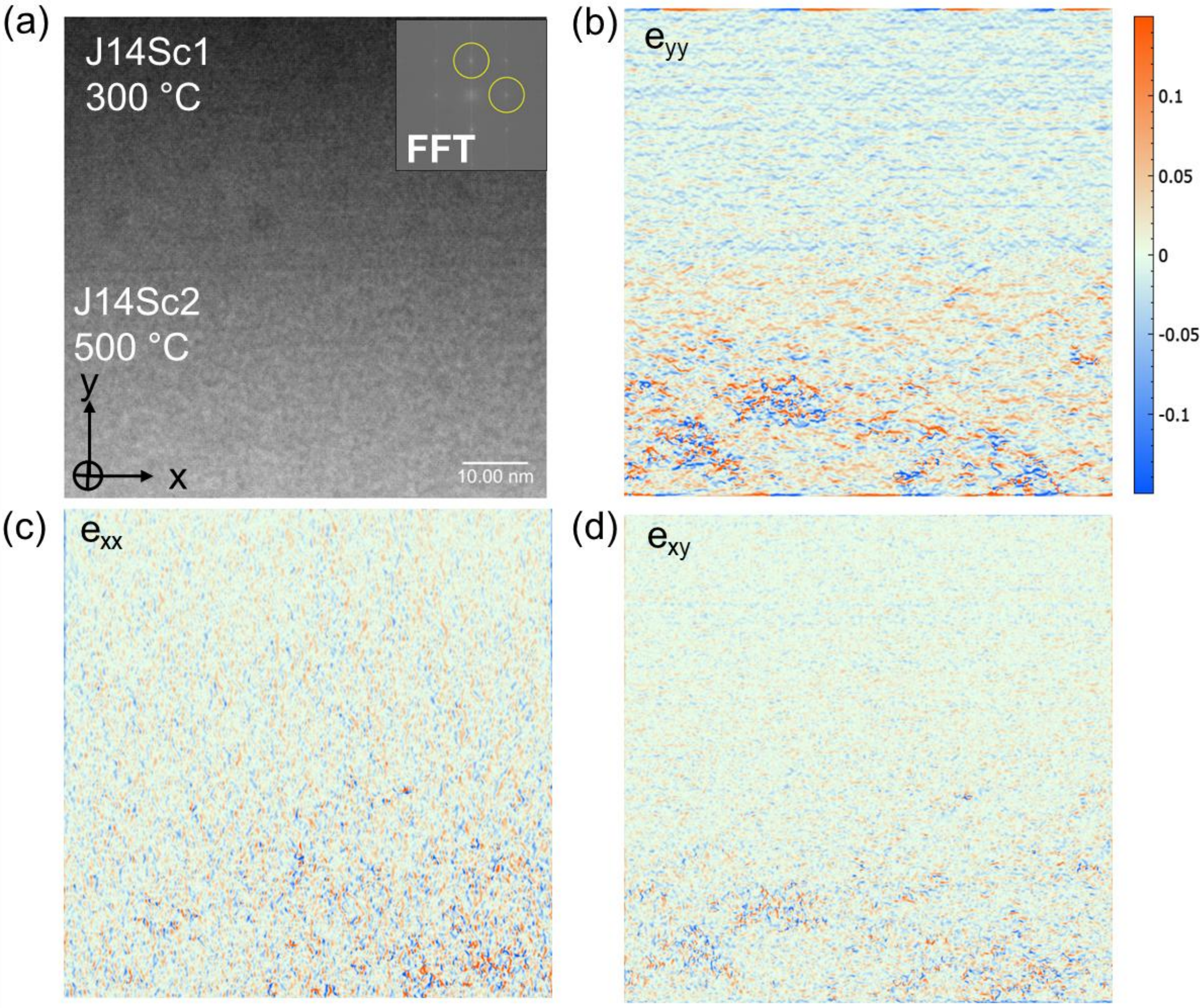

**Figure S7. Strain mapping using geometric phase analysis (GPA) reveals increased out-of-plane tensile strain in films grown at 500 °C.** (a) HAADF-STEM image with an inset showing the FFT and selected reflections used to generate the strain maps (b–d) via GPA.

**S8-S10: Local nuances in electronic states using monochromated STEM-EELS:**

To investigate the variation in electronic states of cations in the HEO thin film, an EELS scan ranging from approximately 510 eV to 980 eV is performed across the entire cross-section of the stacked thin films, as shown in **Figure S8**. The colors of EELS edges correspond to the colors of the probe positions. This energy range enables simultaneous probing of the O K edge, Co L edge, Ni L edge, and Cu L edge. The other cations in the HEO system, Mg and Zn, are consistently present in the 2+ state and are therefore not analyzed.

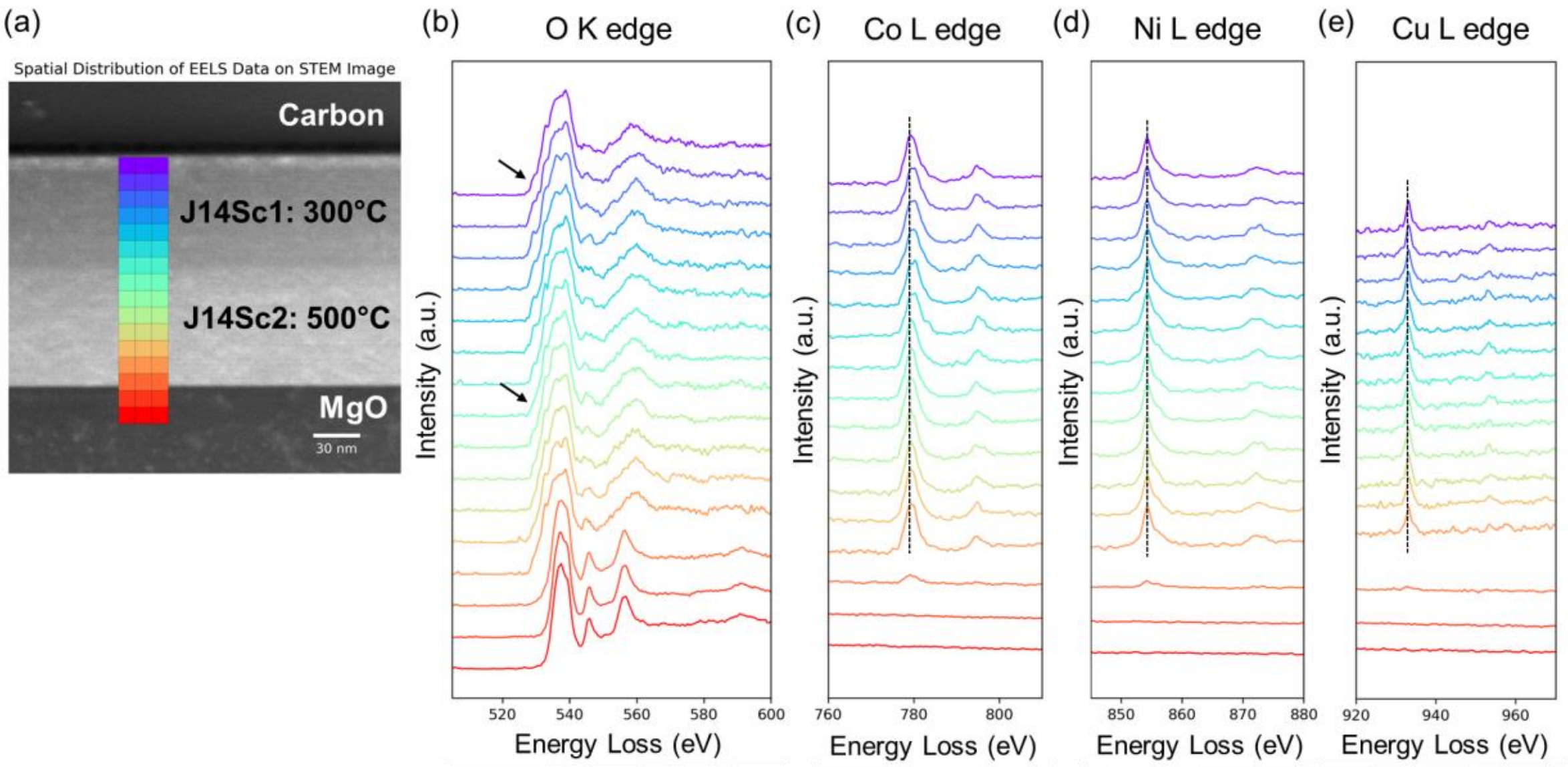

**Figure S8. EELS spectrum across the thin film stack reveals changes in pre-edge in O K edge.** (a) STEM image with each color corresponding to regions from which EELS spectra were simultaneously acquired. (b–e) EELS spectra from the same regions showing the (b) O K-edge, (c) Co L-edge, (d) Ni L-edge, and (e) Cu L-edge.

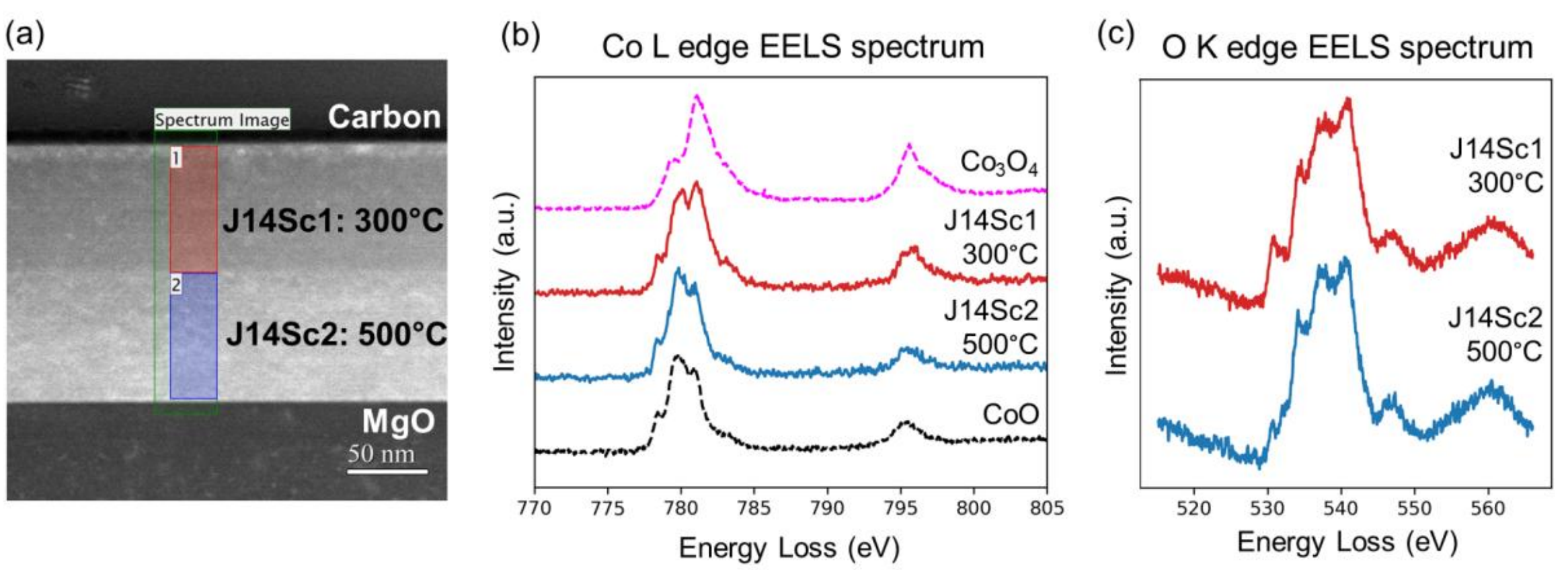

**Figure S9. Summed EELS spectrum of Co L edge and O K edge in J14Sc grown at 300 °C and 500 °C.** (a) STEM image with each color corresponding to regions from which EELS

spectra were summed and plotted in **Figure 4c** (also shown in **Figure 9b**). (b) Same plot as **Figure 4c** for comparison, where average Co L-edge spectra from J14Sc1 and J14Sc2 are plotted alongside $Co^{2+}$ (CoO) and $Co^{2+/3+}$ ($Co_3O_4$) reference EELS data.[1] (c) Summed O K-edge spectra from each thin film, showing a subtle change in the pre-edge peak near 530.8 eV.

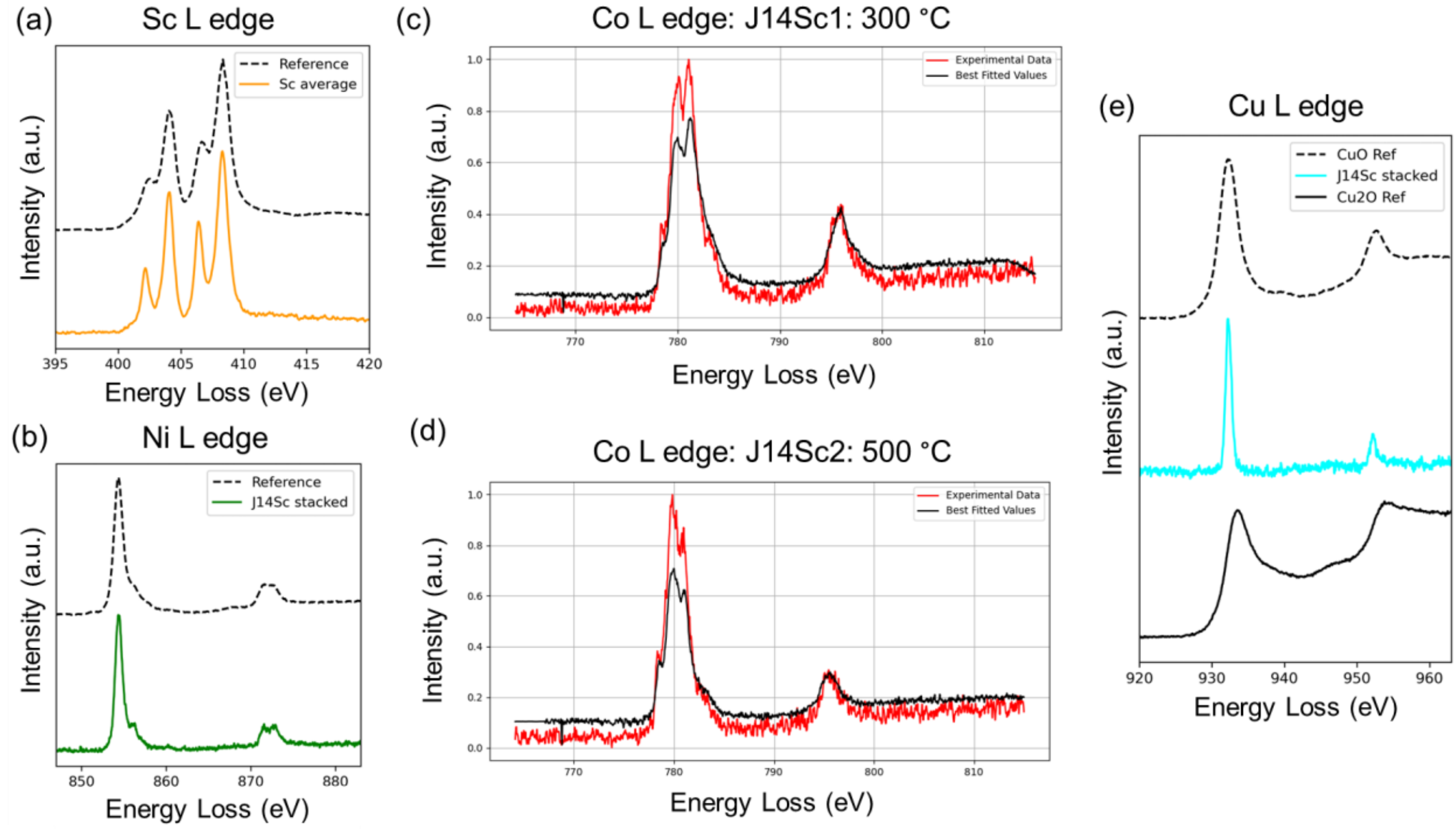

**Figure S10. Monochromated EELS edges of Sc, Ni, and Cu reveal valences of 3+, 2+, and 2+, respectively.** (a–c) Monochromated EELS spectra of (a) Sc L-edge, and (b) Ni L-edge acquired across the entire stacked thin film. (c–d) Least-squares fitting of $Co^{2+}$ and mixed $Co^{2+}/Co^{3+}$ reference spectra in the J14Sc1 and J14Sc2 thin films, respectively. (e) Cu L-edge acquired across the entire stacked thin film plotted with $Cu^{2+}$ (CuO) and $Cu^{1+}$ ($Cu_2O$) references.[1] The Cu $L_3$ peak is symmetric similar to $Cu^{2+}$ reference.

**S11-S12: Structural investigation of stacked thin films:**

In the **Figure 4 (a,b)** unsupervised machine learning algorithm, two clusters were used, as the elbow point from scree plot is around 2, as shown in **Figure S11a**. The HAADF STEM image and corresponding intensity profile reveals a continuous increase in Z-contrast from J14Sc1 to J14Sc2, consistent with the presence of more cation vacancies in J14Sc1, as suggted by the EDS results in **Figure 4d**. We also imaged the same region at different collection angles to investigate the structural changes within the thin film. The strain contrast highlights local strain variations, despite the entire thin film maintaining rock salt structure.

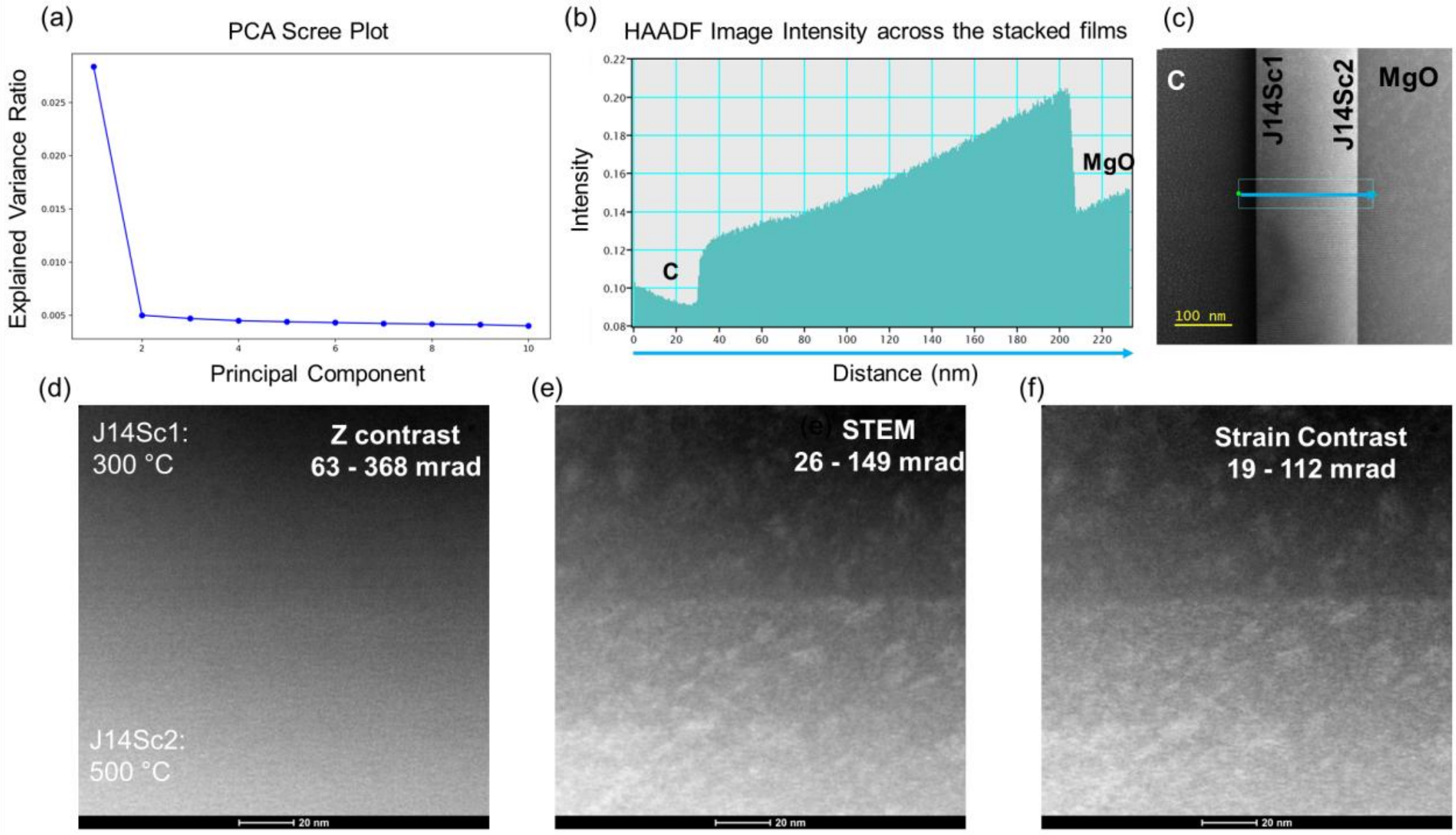

**Figure S11. Structural investigation of stacked thin films.** (a) Scree plot of the sliding window FFT analysis. (b) Intensity profile of the HAADF STEM image shown in (c), indicating a continuous increase in Z contrast from J14Sc1 to J14Sc2. (c) HAADF STEM image with the region used for the summed intensity. (d-e) STEM images acquired at different collection angles, transitioning from Z-contrast to strain contrast, respectively. Local distortions in J14Sc grown at 500 °C can be seen predominantly in strain contrast image, such signatures are visible as subtle streaking in the CBED pattern in **Figure S3c**, oriented perpendicular to the distortions seen in the corresponding real-space region in the VBF image **Figure S3e**.

**S12: Chemical distribution and thickness variation:**

The EDS intensity shows an increasing trend for all elements, including oxygen, and a similar trend is observed in the ADF intensity. The EELS thickness-to-mean free path ratio (t/λ) map reveals local variations across the sample, which could indicate changes in sample thickness and possibly variations in the mean free path due to differences in the chemical environment. To account for the effect of thickness variation, we analyzed the cation intensities relative to oxygen in **Figure 4d**.

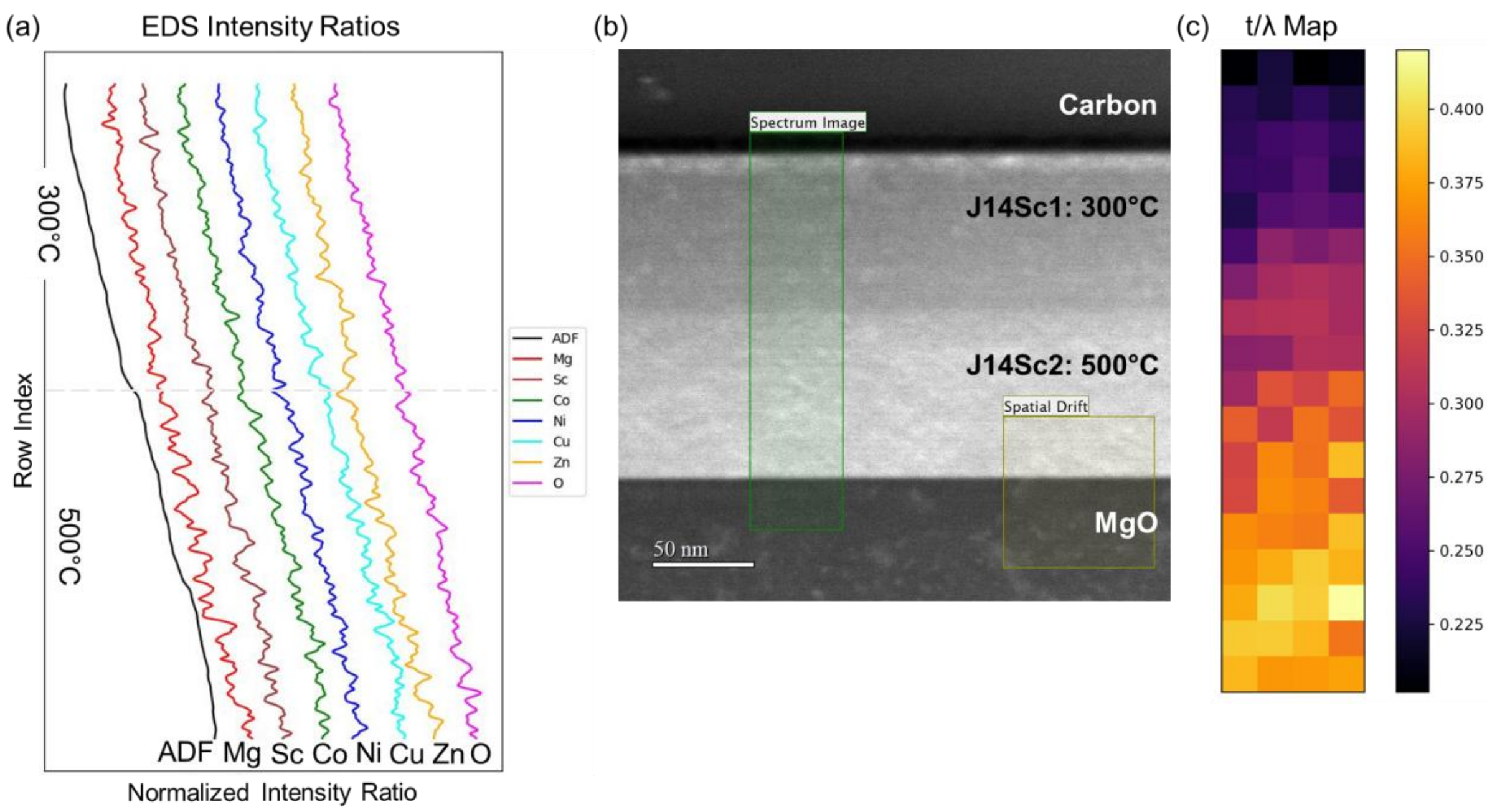

**Figure S12. Chemical distribution and thickness map.** (a) EDS intensity plot of all elements, including oxygen. (b) STEM image acquired during EELS acquisition with the spectrum region indicated by green box, corresponding to the EELS thickness/mean free path map shown in (c). The EELS data was collected from a slightly different region of the same sample compared to the EDS analysis.

**References:**

[1] P. Ewels, T. Sikora, V. Serin, C. P. Ewels, L. Lajaunie, *Microsc. Microanal.* **2016**, *22*, 717.